\documentclass[12pt,a4paper]{report}

\usepackage{geometry}
\usepackage{silence}
\usepackage{amsmath}
\usepackage{url}
\usepackage{graphicx}
\usepackage{pdfpages} 
\usepackage[printonlyused]{acronym}
\usepackage{setspace}
\usepackage{amsthm}
\usepackage{algorithmic}
\usepackage{fancybox}
\usepackage{rotating}
\usepackage[utf8]{inputenc}
\usepackage[lined,boxed,commentsnumbered]{algorithm2e}
\usepackage{amsmath}
\usepackage{ctable}
\usepackage{multirow}
\usepackage[caption=false]{subfig}
\usepackage{rotating}
\usepackage{enumerate}
\usepackage{balance}
\usepackage{breakcites}
\usepackage[hidelinks]{hyperref}

\usepackage{nomencl}
\makenomenclature

\makeindex

\theoremstyle{definition}
 \numberwithin{definition}{chapter} 
 \numberwithin{example}{chapter}

\newcommand{\bllac}{BLL$_{AC}$}
\newcommand{\ourapproach}{BLL$_{AC}$+MP$_r$}
\newcommand{\pred}{BLL$_{I,S}$}
\newcommand{\rec}{BLL$_{I,S,C}$}
\newcommand{\rand}{\textit{Random}}
\newcommand{\res}{\textit{CompSci}}
\newcommand{\sone}{\textit{Scenario 1}}
\newcommand{\stwo}{\textit{Scenario 2}}

\begin{document}

\pagenumbering{arabic}
\setcounter{page}{1}

\chapter*{Modeling Activation Processes in Human Memory to Improve Tag Recommendations\\\textit{by Dominik Kowald}}

This thesis was submitted by Dr. Dominik Kowald\footnote{\url{http://www.dominikkowald.info}} to the Institute of Interactive Systems and Data Science of Graz University of Technology in Austria on the 5th of September 2017 for the attainment of the degree “Dr.techn”. The supervisors of this thesis have been Prof. Stefanie Lindstaedt and Ass.Prof. Elisabeth Lex from Graz University of Technology, and the external assessor has been Prof. Tobias Ley from Tallinn University.

In the current enthusiasm around Data Science and Big Data Analytics, it is important to mention that only theory-guided approaches will truly enable us to fully understand why an algorithm works and how specific results can be explained. It was the goal of this dissertation research to follow this path by demonstrating that a recommender system inspired by human memory theory can have a true impact in the field.

\section*{Problem \& Motivation}
Social tagging systems enable users to collaboratively assign freely chosen keywords (i.e., \textit{tags}) to resources (e.g., Web links, scientific publications, music, images, movies, etc.). These tags can then be used for not only searching, navigating, organizing and finding content but also serendipitous browsing. Therefore, social tags have become an essential instrument of Web 2.0 (i.e., the social Web) to assist users during these activities. Another advantage of social tags is that users can freely choose them for annotating their bookmarked resources. However, this also means that these users have to create a set of descriptive tags on their own, which can be a very demanding task.

As a solution, tag recommendation algorithms have been proposed, which suggest a set of tags for a given user and a given resource. These tag suggestions are typically calculated based on previously used tags and/or the content of resources. Thus, tag recommendation algorithms aim to help not only the individual to find appropriate tags but also the collective to consolidate the shared tag vocabulary with the aim to reach semantic stability and implicit consensus. Furthermore, it was shown that personalized tag recommendations can increase the indexing quality of resources, which makes it easier for users to understand the information content of an indexed resource solely based on its assigned tags.

Another strand of research on the underlying cognitive mechanisms of social tagging has shown that the way users choose tags for annotating resources corresponds to processes and structures in human memory. In this respect, a prominent example is the activation equation of the cognitive architecture ACT-R, which formalizes activation processes in human memory. Specifically, the activation equation determines the activation level (i.e., probability) that a specific memory unit (e.g., a word or tag) will be needed (i.e., activated) in a specific context.

However, while current state-of-the-art tag recommendation approaches perform reasonably well in terms of recommendation accuracy, most of them are designed in a purely data-driven way. Consequently, they are based on either simply counting tag frequencies or computationally expensive calculation steps (e.g., calculating user similarities, modeling topics  or factorizing the features of resources). Hence, these approaches typically ignore the above mentioned insights originating from cognitive research on how humans access information, such as words or tags, in their memory. This is contrary to the assumption that tag recommendation algorithms should attempt to mimic the user's tagging behavior.

Thus, the main problem this thesis aims to tackle is to show that a cognitive-inspired approach, which is build upon activation processes in human memory, can improve the state-of-the-art of tag recommendations. Furthermore, such an approach should help to better understand the underlying cognitive processes of social tagging and tag recommendations. The main idea of this thesis was presented at WWW'2015 \cite{P1}.

\section*{Approach \& Methods}
Consider a user retrieving a unit from her memory, such as a tag that she has used previously. To derive its usefulness in the current context, the activation level $A_i$ of this memory unit $i$ has to be determined. According to the following activation equation, which is part of the declarative module of the cognitive architecture ACT-R, the usefulness of $i$ is given by:
 \begin{equation} \label{eq:A}
		A_i = B_i + \sum_j{W_j \cdot S_{j,i}}
  \end{equation}
The $B_i$ component represents the \textit{base-level} activation and quantifies the general usefulness of a unit $i$ by considering how frequently and recently it has been used in the past. It is given by the base-level learning (BLL) equation:
\begin{equation} \label{eq:bll}
    B_i = ln(\sum\limits_{j = 1}\limits^{n}{t_{j}^{-d}})
  \end{equation}
where $n$ is the frequency of the unit's occurrences and $t_j$ is the recency (i.e., the time in seconds since the $j^{th}$ occurrence of $i$). For example, if a user has applied the two tags ``recognition'' and ``recommender'' with equal frequency but ``recommender'' has dominated the user's recent bookmarks, the equation predicts a higher activation level for ``recommender''. The exponent $d$ accounts for the power-law of forgetting, which means that each unit's activation level caused by the $j^{th}$ occurrence decreases in time according to a power function. $d$ is typically set to $0.5$. \cite{anderson2004integrated}.

The second component of Equation \ref{eq:A} represents the associative component (i.e., the \textit{contextualized priming}) that tunes the base-level activation of the unit $i$ to the current semantic context. The context is given by any contextual element $j$ important in the current situation (e.g., the tags ``memory'' and ``recollection''). Through learned associations, the contextual elements are connected with tag $i$ and can increase $i$'s activation depending on the weight $W_j$ and the strength of association $S_{j,i}$. To simplify matters, the tags associated with a given resource $r$ (due to previous tag assignments of other users) are used as the contextual elements. The weight $W_j$ is derived from the number of times tag $j$ has been assigned to resource $r$, and $S_{j,i}$ is derived from the number of co-occurrences between the tags $i$ and $j$. The theoretical fundamentals behind this approach were presented at WWW'2014 \cite{P3} and in \cite{P4,P5}.

To evaluate this approach, common practice in research on recommender systems was followed to build on an offline evaluation study design. This means that publicly available social tagging datasets (i.e., Flickr, CiteULike, BibSonomy, Delicious, MovieLens, LastFM) were used in order to study the relation between activation processes in human memory and tagging behavior of users on a large scale. Additionally, these data collections were split into training and test sets in order to conduct an offline tag recommender study. For demonstrating the generalizability of the proposed approach, additionally, data collections from Twitter were used. To foster principles of Open Science, especially with respect to the reproducibility of the results, the complete evaluation pipeline was developed as open-source software in form of the \textit{TagRec} framework and put on GitHub\footnote{\url{https://github.com/learning-layers/TagRec}}. The \textit{TagRec} framework was presented at Hypertext'2014 \cite{P6} and UMAP'2017 \cite{P7}.

\section*{Results \& Findings}

\subsection*{\textit{Result 1}: The Influence of Activation Processes in Human Memory on Tag Reuse}
The first main result of this thesis deals with the relation between activation processes in human memory and the tag reuse behavior of users in social tagging systems. According to \cite{anderson2004integrated}, the activation of a memory unit (e.g., a tag) should depend on at least two variables: (i) the general usefulness of this memory unit given by past usage frequency and recency, and (ii) its usefulness in the current semantic context. Therefore, it is the aim of \textit{Result 1} to test if this is also applicable for social tagging settings.

In order to quantify the influence of usage frequency, recency and semantic context on the reuse of tags, the tag assignments of the first $n - 1$ bookmarks (i.e., reflecting the past) of a user $u$ were compared with the tag assignments of $u$'s $n^{th}$ bookmark (i.e., reflecting the future). Within this evaluation, it is also validated if the factor of recency can be better modeled via a power or an exponential function.

\begin{figure}[ht]
   \centering
	 \subfloat[Tag frequency]{ 
      \includegraphics[width=0.33\textwidth]{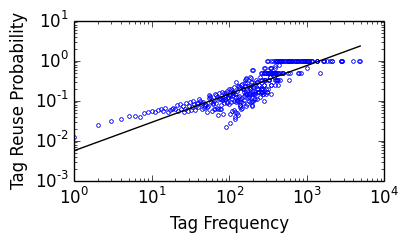}%
   }
	 \subfloat[Tag recency]{ 
      \includegraphics[width=0.33\textwidth]{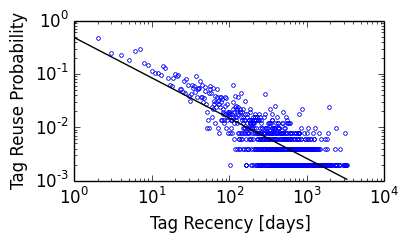}%
   }
		\subfloat[Semantic Context]{ 
      \includegraphics[width=0.33\textwidth]{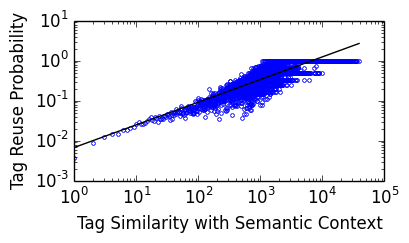}%
   }
   \caption{The influence of activation processes in human memory (i.e., usage frequency, recency and semantic context) on tag reuse in CiteULike (\textit{Result 1}).}
	 \label{fig:rq1_frequency}
\end{figure}

In this respect, the results presented in Figure \ref{fig:rq1_frequency} lead to the following four findings: (i) the \textit{more frequently} a tag was used in the past, the higher its probability of being reused, (ii) the \textit{more recently} a tag was used in the past, the higher its probability of being reused, (iii) the \textit{more similar} a tag is to tags in the \textit{current semantic context}, the higher its probability of being reused, and (iv) the effect of \textit{recency on the reuse probability of tags} is more likely to follow a \textit{power-law distribution} than an exponential one. These findings were presented at Hypertext'2016 \cite{P2}.

\subsection*{\textit{Result 2}: Evaluating a Cognitive-Inspired Algorithm for Tag Recommendations}
Based on the outcomes of \textit{Result 1}, a novel cognitive-inspired approach for predicting the reuse of tags (i.e., \bllac{}) was implemented and evaluated. Therefore, the prediction accuracy performance of \bllac{} was measured by means of the F1@5 and nDCG@10 metrics, and compared to current state-of-the-art algorithms, namely MostPopular (MP$_{u, r}$), Collaborative Filtering (CF), Latent Dirichlet Allocation (LDA), Temporal Tag Patterns (GIRPTM), Pairwise Interaction Tensor Factorization (PITF) and FolkRank (FR). This comparison is given in Table \ref{tab:rq3_results} and shows that \bllac{} outperforms these approaches in four of the six datasets (i.e., Flickr, CiteULike, BibSonomy and Delicious). These four datasets represent narrow folksonomies, in which typically only one or a few users annotate the same resource. In LastFM and MovieLens, however, PITF and FR provide the best results, which shows the importance of incorporating also social influences for tag recommendations by means of imitating popular tags of other users. The main reason for this is that LastFM and MovieLens represent broad folksonomies, in which typically a lot of users annotate the same resource and thus, tag imitation becomes much more important.

\begin{table}[ht]
  \setlength{\tabcolsep}{2.0pt}	
  \centering
    \begin{tabular}{l|l|cccccc|c|c}
    \specialrule{.2em}{.1em}{.1em}
	Dataset	& Metric	& MP$_{u, r}$		& CF	& LDA					& GIRPTM 		& PITF				& FR	& \bllac{}  & \ourapproach{} 			\\\hline 
		\multirow{2}{*}{\centering{\centering{Flickr}}}																	                
 & F1@5				& .371					& .453			& .178						& .455				& .350				& .365						& .470					& \textbf{.470} \\
 & nDCG@10				& .569					& .666			& .280					& .686				& .535				& .561						& .711		& \textbf{.711} \\\hline
		\multirow{2}{*}{\centering{\centering{CiteULike}}}																				      
 & F1@5			& .249					& .231		  & .089	 		  & .262			 & .178				& .250					& .259					& \textbf{.273}			\\
& nDCG@10		& .392					& .359		  & .138	  	  & .420			& .294				& .392						& .422					& \textbf{.438 }\\\hline											
		\multirow{2}{*}{\centering{\centering{BibSonomy}}}																			        
& F1@5			& .281					& .260			& .145				& .291			& .215						& .279					& .279					& \textbf{.298}		\\
& nDCG@10			& .407					& .369			& .219				& .425			& .327						& .408					& .409					& \textbf{.434}		\\\hline
	\multirow{2}{*}{\centering{\centering{Delicious}}}																		          
& F1@5		& .238					& .243			& .182				 	& .261			 & .199						& .196				& .243					& \textbf{.283}				\\
& nDCG@10			& .358					& .356			& .271			& .393						  & .302						& .292					& .374					& \textbf{.431}		\\\hline
	\multirow{2}{*}{\centering{\centering{LastFM}}}														                      
& F1@5		& .258					& .226			& .258				& .263							& .276						& .270					& .251					& \textbf{.283}			\\
& nDCG@10		& .386					& .317			& .388					& .397				& .414						& .399							& .375					& \textbf{.425}			\\\hline
	\multirow{2}{*}{\centering{\centering{MovieLens}}}														                  
& F1@5	& .153					& .124			& .141				& .159		& .156						& .153					& .086					& \textbf{.160}							\\
& nDCG@10		& .328					& .254			& .296			& .326				& .324						& .319							& .203					& \textbf{.338}			\\
		\specialrule{.2em}{.1em}{.1em}								
    \end{tabular}
    \caption{Recommendation accuracy evaluation results by means of the F1@5 and nDCG@10 metrics. The cognitive-inspired approach \ourapproach{} provides the best results in all evaluation settings (\textit{Result 2}).}	
  \label{tab:rq3_results}
\end{table}

To address this, a hybrid recommendation algorithm was implemented and evaluated. In this respect, the contributions are twofold. Firstly, it was shown that \bllac{} can be expanded with tag imitation processes by means of popular tags of other users in order to realize \ourapproach{}. Secondly, according to the evaluation results presented in Table \ref{tab:rq3_results}, \ourapproach{} now also outperforms PITF and FR in all six datasets (i.e., narrow and broad folksonomies). These findings were presented at RecSys'2015 \cite{P8}.

\subsection*{\textit{Result 3}: Utilizing the Approach for Hashtag Recommendations in Twitter}
In order to demonstrate the generalizability of this approach, the third main result of this thesis was the design, implementation and evaluation procedure of a cognitive-inspired approach for hashtag recommendations in Twitter. Based on the findings of \textit{Result 1} and \textit{Result 2}, this algorithm utilizes activation processes in human memory to account for temporal effects on individual hashtag reuse (i.e., reusing own hashtags) and social hashtag reuse (i.e., reusing hashtags, which has been previously used by a followee). Therefore, an analysis of hashtag usage types in two empirical networks (i.e., \res{} and \rand{} datasets) crawled from Twitter was conducted, which revealed that between 66\% and 81\% of hashtag assignments can be explained by past individual and social hashtag usage. By analyzing the timestamps of these hashtag assignments, it was further shown that temporal effects play an important role for both individual and social reuse of hashtags and that a power function provides a better fit to model this time-dependent decay than an exponential function.

\begin{figure}[ht]
   \centering
	 \captionsetup[subfigure]{justification=centering}
	 \subfloat[][\sone{}: Hashtag recommendation\\w/o current tweet\\\res{} dataset]{ 
      \includegraphics[width=0.24\textwidth]{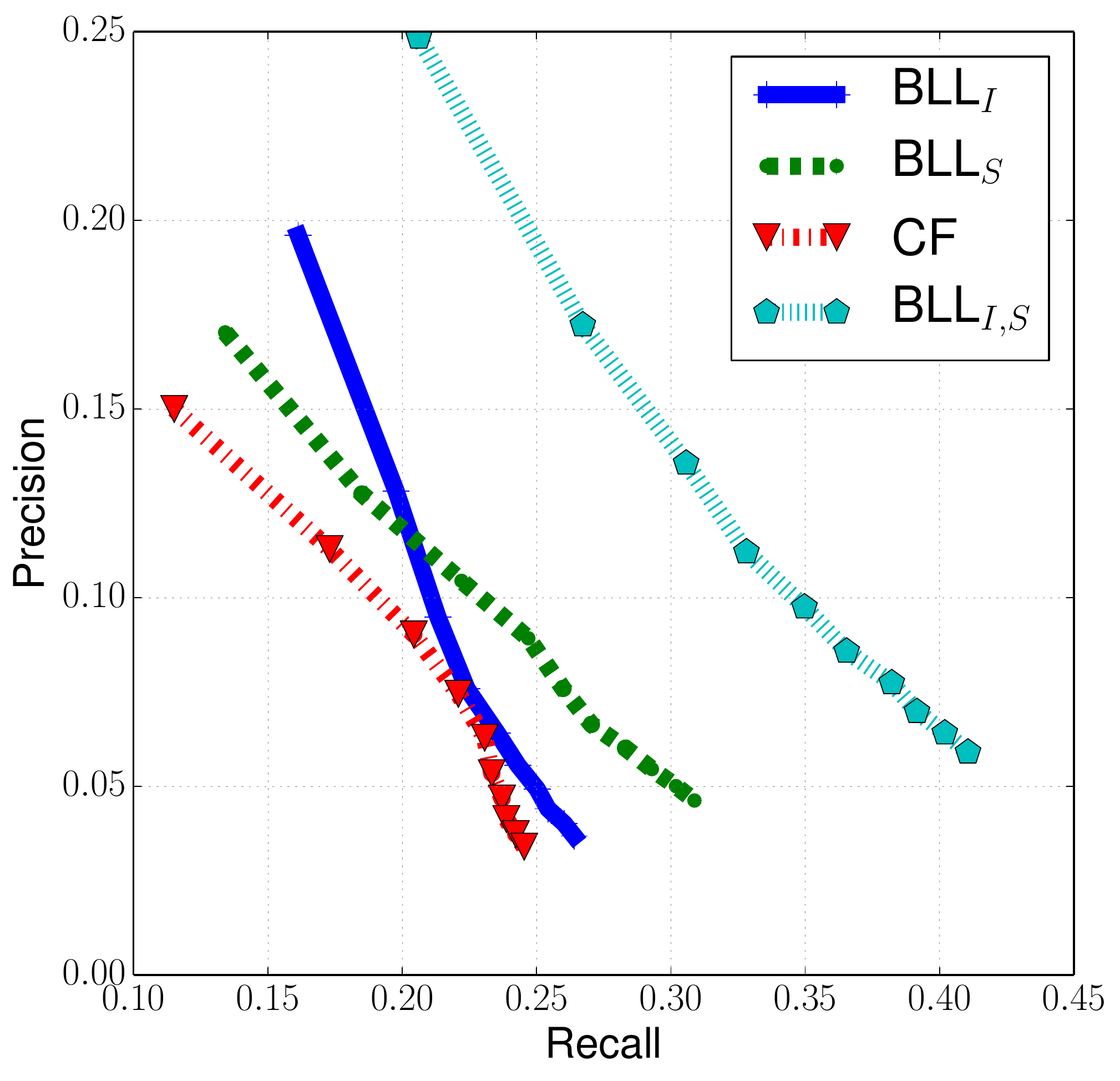} 
   }
	 \subfloat[][\sone{}: Hashtag recommendation\\w/o current tweet\\\rand{} dataset]{ 
      \includegraphics[width=0.24\textwidth]{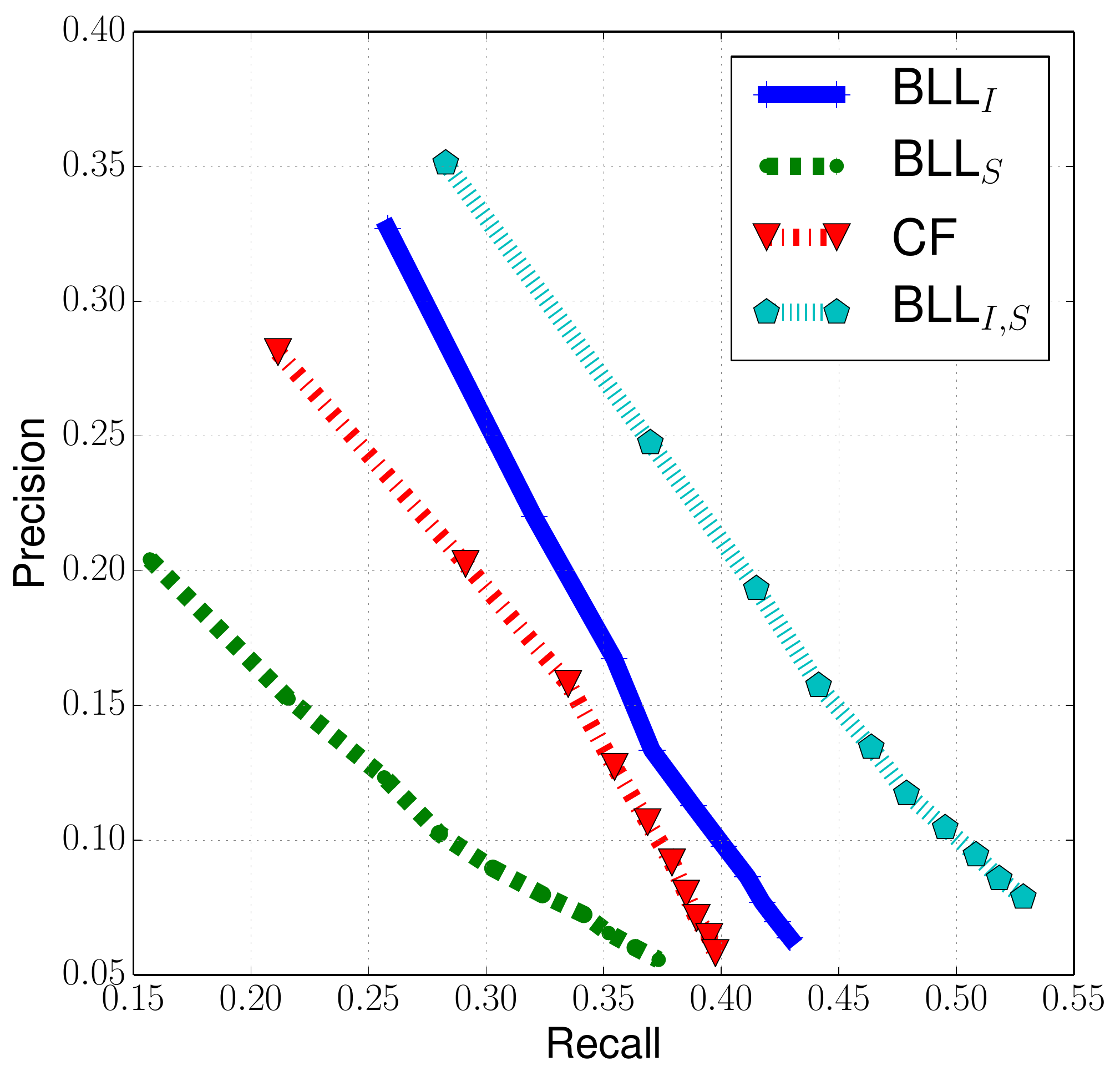} 
   }
	 \subfloat[][\stwo{}: Hashtag recommendation\\w/ current tweet\\\res{} dataset]{ 
      \includegraphics[width=0.24\textwidth]{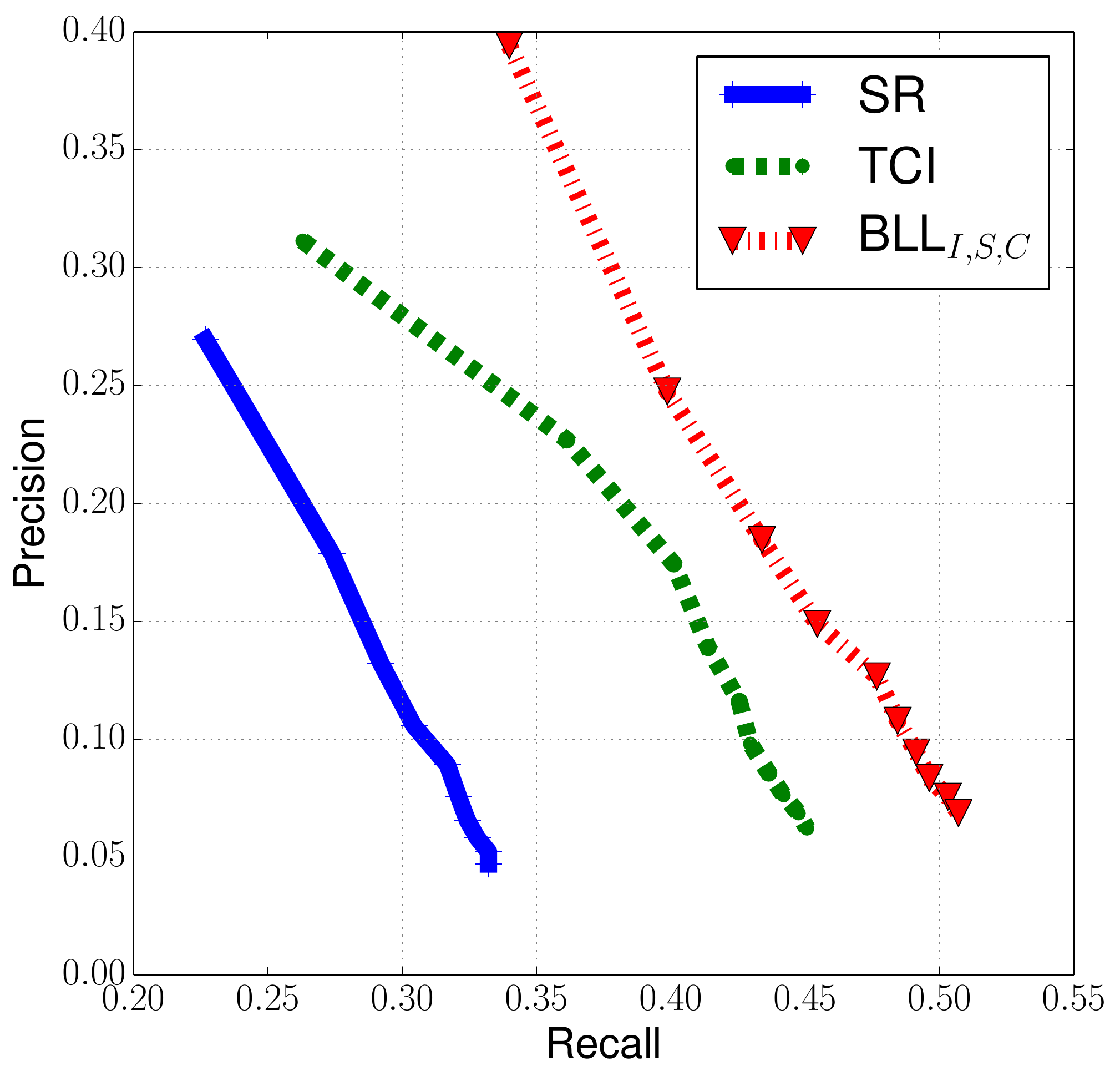} 
   }
	 \subfloat[][\stwo{}: Hashtag recommendation\\w/ current tweet\\\rand{} dataset]{ 
      \includegraphics[width=0.24\textwidth]{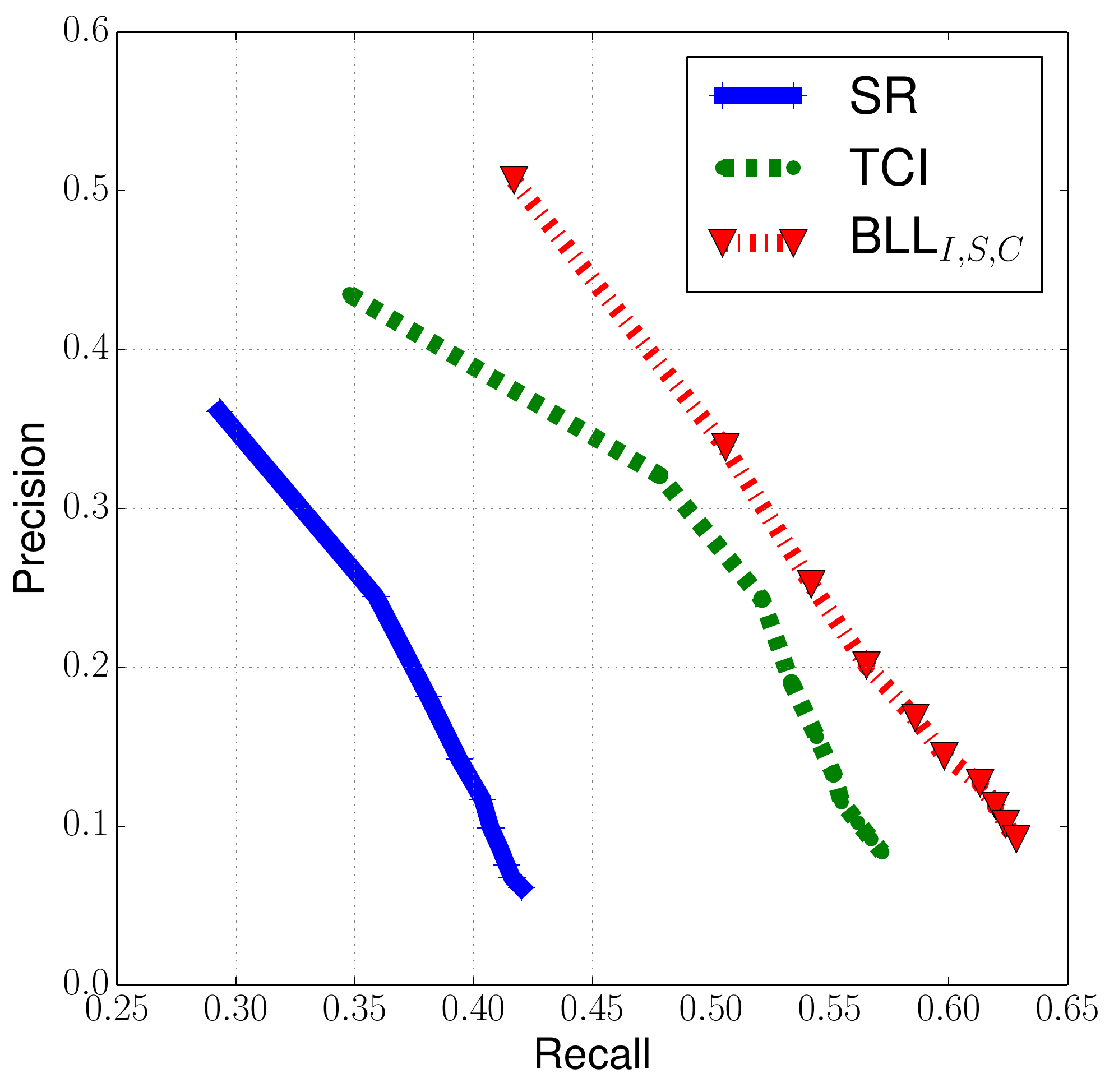} 
   }
   \caption{Recommendation accuracy results by means of Precision / Recall plots for hashtag recommendations. \pred{} and \rec{} outperform the current state-of-the-art in both evaluation settings and in both datasets (\textit{Result 3}).}
	 \label{fig:results}
\end{figure}

Based on this, the activation equation of ACT-R was utilized to develop \pred{} and \rec{}, two algorithms for recommending hashtags. Whereas \pred{} aims to recommend hashtags without incorporating the current tweet (i.e., \sone{}), \rec{} also utilizes the content of the current tweet using the TF-IDF statistic (i.e., \stwo{}). Both algorithms were compared to state-of-the-art hashtag recommendation algorithms, namely BLL$_{individiual}$ (BLL$_I$), BLL$_{social}$ (BLL$_S$), Collaborative Filtering (CF), SimilarityRank (SR) and TemporalCombInt (TCI). This evaluation is shown in Figure \ref{fig:results}, which leads two four main findings: (i) a substantial amount of hashtag assignments in Twitter can be explained by \textit{past individual and social hashtag usage}, (ii) \textit{temporal effects} have an important influence on both \textit{individual as well as social hashtag reuse}, (iii) \textit{a power function} is better suited to model this \textit{time-dependent decay} than an exponential one, and (iv) \pred{} outperforms related algorithms in the first evaluation scenario (i.e., \textit{without the current tweet}) and \rec{} in the second one (i.e., \textit{with the current tweet}).

These findings were presented at WWW'2017 \cite{P10} and show that activation processes in human memory can be utilized for the recommendation of hashtags in Twitter. This further demonstrates that the cognitive-inspired tag recommendation approach proposed in this thesis can be generalized for related use cases in the area of tag-based recommender systems.

\section*{Impact \& Future Work}
This thesis has modeled activation processes in human memory in order to improve tag recommendations. Therefore, the relation between activation processes in human memory and the reuse of tags in social tagging system has been investigated and a novel tag recommendation algorithm termed \ourapproach{} has been proposed based on the activation equation of the cognitive architecture ACT-R. This algorithm has been evaluated in six real-work folksonomy datasets to demonstrate that a cognitive-inspired approach is able to outperform state-of-the-art tag recommendation methods that ignore these insights from cognitive science. Finally, the algorithm was adapted and extended for recommending hashtags in Twitter, which shows that activation processes in human memory can be utilized for related use cases in the area of tag-based recommender systems.

The impact of this dissertation research can be found in the areas of (i) recommender systems, (ii) social tagging, and (iii) tag recommendation evaluation. With respect to the research area of recommender systems, this thesis showed that principles of human cognition can be utilized to improve tag recommendations. By taking into account how humans access information in their memory, cognitive-inspired recommendation strategies were developed as an alternative to data-driven methods such as Collaborative Filtering or Matrix Factorization. These findings had an impact on the research area of recommender systems by motivating the development of related cognitive-inspired recommendation algorithms (e.g., \cite{P1,P3}).

Furthermore, this thesis contributed to the large body of research that analyzes interactions in social tagging systems. Specifically, the relation between activation processes in human memory and the reuse of tags was investigated, which resulted in a set of factors that influence the reuse of tags. Thus, three of these factors are (i) past usage frequency, (ii) past usage recency, and (iii) the current semantic context cues. Especially, the second factor (i.e., time / recency) has led to further interesting investigations such as the study of temporal effects on hashtag reuse in Twitter (e.g, \cite{P2,P10}). Finally, this thesis provided the to-date most extensive evaluation of tag recommendation algorithms. This included the evaluation of twenty tag recommendation algorithms on six datasets by using ten evaluation metrics. This large-scale study provided the research community with a performance overview of state-of-the-art tag recommendation algorithms in various evaluation settings. Apart from that, all algorithms and evaluation procedures used in this thesis were provided as open-source software, which greatly enhanced the reproducibility of the presented results (e.g., \cite{P7,P8,P9}).

This thesis showed that activation processes in human memory can be utilized for both tag recommendations in social tagging systems and hashtag recommendations in Twitter. These findings open up possible research strands for future work, such as the design of cognitive-inspired recommender systems \cite{P11,P12,P13,lacic2014towards,lacic2014socrecm,lacic2015utilizing}. One example of such cognitive-inspired resource recommendation algorithm was presented at WWW'2015 \cite{seitlinger2015attention} and in \cite{kopeinik2016improving}.

Taken together, the author of this thesis believes, that general future research of recommender systems should focus on a hybrid combination of (i) cognitive-inspired, and (ii) data-driven (e.g., machine learning-based) approaches. This way, the strengths of both types of algorithms can be combined in order to adapt to the given data (i.e., large-scale versus small-scale settings). This thesis already provided a first step into this research direction.

\end{document}